\documentclass{article}
\usepackage{spconf,amsmath,graphicx,hyperref}

\usepackage{amssymb}
\usepackage{booktabs}
\usepackage{multirow}
\usepackage{array}
\usepackage[table]{xcolor}
\usepackage{xurl}

\title{Augmenting Large Audio-Language Models with Frame-Level Grounding for Fine-Grained Temporal Perception}
\name{
Yanfeng Shi$^{1}$, Yan Song$^{1}$, Junhui Li$^{1}$, Tinggan Huang$^{1}$, Wu Guo$^{1}$,
Haoyu Song$^{2}$, Ian McLoughlin$^{2}$
}

\address{
$^{1}$National Engineering Research Center of Speech and Language Information Processing,\\
University of Science and Technology of China, Hefei, China\\
$^{2}$ICT Cluster, Singapore Institute of Technology, Singapore
}

\begin{document}
\ninept
\maketitle
\begin{abstract}
Large Audio-Language Models (LALMs) have substantially advanced general audio understanding, yet they remain limited in fine-grained temporal perception, particularly in precise event localization.
Existing approaches primarily post-train LALMs to predict event boundaries as timestamp tokens.
However, this generative formulation lacks explicit correspondence between the timestamp predictions and fine-grained acoustic evidence, limiting the precision and reliability of temporal localization.
To address this issue, we augment the LALM with a dedicated frame-level grounding model while leveraging its semantic modeling capability to represent the event query.
Specifically, the frozen LALM encodes the event query with audio as context, and the grounding model combines these query representations with fine-grained audio features to localize the target event at the frame level.
Extensive experiments across diverse temporal grounding benchmarks demonstrate strong and consistent improvements over existing methods.
Further evaluation shows that the grounding model can provide temporal evidence to support downstream reasoning.
\footnote{%
  \urlstyle{same}%
  \def\UrlBreakPenalty{0}%
  \def\UrlBigBreakPenalty{0}%
  Examples of temporal grounding and reasoning are available at
  \url{https://saber5203.github.io/frame-level-grounding/}.%
}
\end{abstract}
\section{Introduction}
\label{sec:intro}

Large Audio-Language Models (LALMs) have significantly advanced general audio understanding by integrating acoustic perception with the semantic knowledge and reasoning capabilities of large language models~\cite{ltu,salmonn,audio_flamingo}.
Through multimodal pretraining and instruction tuning, LALMs acquire the ability to interpret diverse acoustic content, including speech, environmental sounds, and music, guided by natural language instructions.
However, existing LALMs remain limited in fine-grained temporal perception, particularly in the precise localization of audio events~\cite{timeaudio}.
This limitation restricts their applicability to temporal tasks such as audio grounding~\cite{audiogrounding} and can further undermine downstream reasoning that depends on reliable temporal evidence, such as determining which sound event lasts the longest.

Recent work has sought to address this limitation by enhancing the temporal localization capabilities of LALMs~\cite{eodn,spotsound,timepro-rl}.
Temporal information is encoded using special timestamp tokens~\cite{timeaudio} or incorporated into audio feature sequences through temporal markers~\cite{moss_audio}, offering temporal references for localization.
Meanwhile, fine-grained annotations provide supervision for temporal grounding~\cite{spotsound}, while reinforcement learning further refines localization precision~\cite{timepro-rl}.
Collectively, these efforts have substantially improved the temporal grounding performance of LALMs.

Notably, existing methods largely follow a common paradigm in which the LALM is post-trained for fine-grained temporal localization, with event boundaries predicted as timestamp tokens through its text generation interface.
LALMs are well suited to interpreting natural language queries through their semantic representations, whereas precise temporal localization requires resolving event activity based on fine-grained acoustic evidence.
Under the generative formulation, the correspondence between temporal predictions and fine-grained acoustic cues is learned implicitly, which can compromise the precision and reliability of temporal localization.
By contrast, frame-level prediction directly models event activity over the audio timeline and has long been adopted in sound event detection~\cite{sed_tutorial} for precise temporal localization~\cite{sed_crnn,atst_sed}.
However, how to augment LALMs through frame-level prediction to improve their fine-grained temporal perception while leveraging their semantic modeling capability remains underexplored.

Motivated by these observations, we propose a framework that leverages a frozen LALM to represent queries in context and delegates fine-grained temporal localization to a dedicated frame-level grounding model.
Specifically, given the audio and query prompt as input, we extract hidden states from the LALM as query representations for temporal grounding.
The grounding model combines these representations with frame-level audio features from a pretrained audio encoder through adaptive frame-level alignment and fusion, and the resulting sequence is then temporally modeled to predict frame-level localization scores over the audio timeline.
Extensive experiments demonstrate strong performance across diverse temporal grounding benchmarks, supporting the effectiveness of augmenting LALMs with frame-level grounding.
Further, we employ the grounding model as an external tool and observe improvements on downstream temporal reasoning.

\section{Method}
\label{sec:method}

\begin{figure*}[!t]
    \centering
    \includegraphics[width=0.91\textwidth]{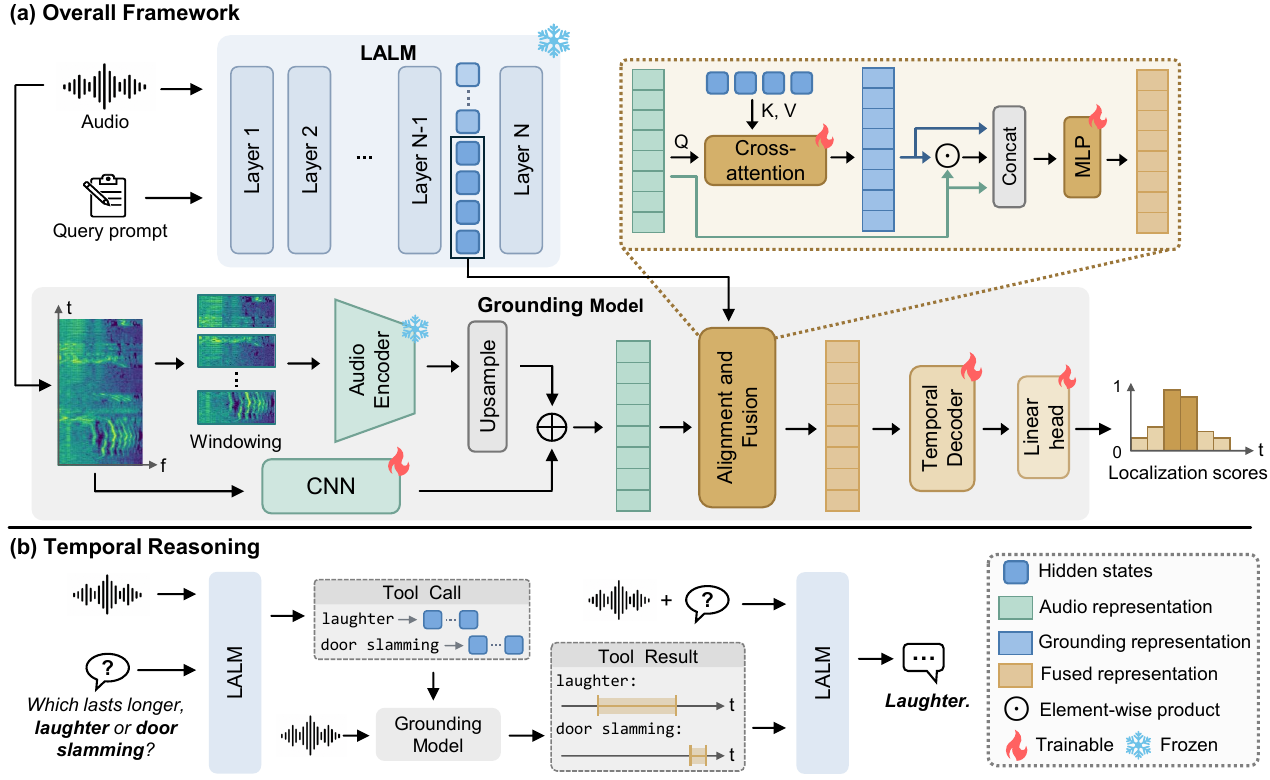}
    \caption{The proposed framework is illustrated in (a), and (b) presents the grounding model as an external tool that provides temporal evidence for LALM reasoning.}
    \label{fig:framework}
\end{figure*}

In this section, we detail the proposed framework illustrated in Fig.~\ref{fig:framework}, including its application to downstream temporal reasoning.

\subsection{LALM Query Representation}
\label{sec:lalm_query}

Rather than adapting the LALM for fine-grained temporal localization, we keep it frozen and leverage its semantic representations for grounding.
The LALM can contextualize the event query with the input audio and localization instruction.
Accordingly, we use its hidden states to condition the subsequent frame-level grounding model.

Specifically, the audio and query prompt are provided as input to the LALM:
``\texttt{<audio> When does "<query>" occur in the audio?}'', where \texttt{<query>} is replaced by the target event description and \texttt{<audio>} by the audio embeddings produced by the LALM's native audio encoder.
We use hidden states from the penultimate layer, which have been reported to encode richer semantic information than those from the final layer~\cite{audio_omni}.
From this sequence, only the hidden states corresponding to the textual tokens are retained.
These states encode the query and localization instruction while incorporating the preceding audio context through self-attention.
We discard the audio-side hidden states because frame-level acoustic information is provided separately by a pretrained audio encoder.
The retained states are then projected to the shared feature space:
\begin{equation}
\widetilde{H}
=
\operatorname{Proj}_{h}(H_{\text{text}})
\in \mathbb{R}^{L\times d},
\end{equation}
where $H_{\text{text}}$ denotes the extracted textual hidden states, $L$ is the number of textual tokens, and $d$ is the shared feature dimension.

\subsection{Frame-Level Grounding}
\label{sec:frame_grounding}

\textbf{Audio Representation}.
Precise temporal localization requires audio features with strong event discrimination and fine temporal structure.
Pretrained audio encoders such as BEATs~\cite{beats} and EAT~\cite{eat} provide strong representations for audio events, and ATST-Frame~\cite{atst_frame} is further tailored to frame-level tasks.
We therefore employ a pretrained audio encoder to obtain frame-level audio representations and keep it frozen during training.

To match the input configuration of the pretrained audio encoder, recordings are divided into non-overlapping 10s windows, with the final window zero-padded to 10s when necessary.
Each window is encoded independently, and the resulting features are concatenated along the temporal dimension to form the encoder sequence for the complete recording as $X_{\mathrm{enc}}=\operatorname{Concat}_{t}\!\left(\mathcal{E}(m_1),\ldots,\mathcal{E}(m_K)\right)$, where $\mathcal{E}$ denotes the audio encoder and $\{m_k\}_{k=1}^{K}$ are the log-Mel windows.
$X_{\mathrm{enc}}$ is then upsampled to the target temporal resolution, while a CNN branch operating on the log-Mel spectrogram supplements the upsampled features with fine-grained acoustic details~\cite{dasm}, resulting in the final audio representation:
\begin{equation}
X
=
\operatorname{Proj}_{x}\!\left(
X_{\mathrm{enc}}^{\uparrow}
+
wX_{\mathrm{cnn}}
\right)
=
\{x_t\}_{t=1}^{T}
\in \mathbb{R}^{T\times d},
\end{equation}
where $X_{\mathrm{enc}}^{\uparrow}$ denotes the upsampled encoder features, $X_{\mathrm{cnn}}$ denotes the CNN features, $w$ is a learnable scalar, and $T$ is the number of frames.

\begin{table*}[!t]
    \centering
    \caption{
    Main temporal grounding results.
    R@.5 and R@.7 denote recall computed at IoU thresholds of 0.5 and 0.7, respectively, and mIoU denotes mean IoU.
    Best and second-best results are highlighted in bold and underlined.
    }
    \label{tab:main_results}
    \renewcommand\arraystretch{1.22}
    \addtolength{\tabcolsep}{-1pt}
    \resizebox{\textwidth}{!}{%
    \begin{tabular}{@{}l ccc|ccc|ccc|ccc|ccc@{}}
        \toprule
        \multirow{2}{*}{\textbf{Method}}
        & \multicolumn{3}{c|}{\textbf{AudioGrounding}}
        & \multicolumn{3}{c|}{\textbf{DESED}}
        & \multicolumn{3}{c|}{\textbf{UnAV-100 subset}}
        & \multicolumn{3}{c|}{\textbf{TACOS}}
        & \multicolumn{3}{c}{\textbf{Clotho-Moment}} \\
        \cmidrule(lr){2-4}
        \cmidrule(lr){5-7}
        \cmidrule(lr){8-10}
        \cmidrule(lr){11-13}
        \cmidrule(lr){14-16}
        & R@.5 & R@.7 & mIoU
        & R@.5 & R@.7 & mIoU
        & R@.5 & R@.7 & mIoU
        & R@.5 & R@.7 & mIoU
        & R@.5 & R@.7 & mIoU \\
        \midrule

        AM-DETR~\cite{am_detr}
        & 15.4 & 5.7 & 29.6
        & 20.3 & 9.6 & 33.7
        & 57.0 & 37.0 & 50.3
        & 7.8 & 3.4 & 21.6
        & 85.8 & 80.3 & 79.5 \\

        FineLAP~\cite{finelap}
        & 65.8 & 48.0 & 61.5
        & 74.2 & 54.3 & 68.4
        & 43.0 & 26.0 & 44.4
        & 36.2 & 27.2 & 39.3
        & 42.6 & 30.7 & 43.3 \\

        DASM~\cite{dasm}
        & 68.8 & 48.8 & 64.0
        & 81.0 & 64.7 & 73.9
        & 32.0 & 11.0 & 39.1
        & 42.4 & 32.0 & 46.7
        & 18.9 & 5.9 & 32.8 \\

        MOSS-Audio~\cite{moss_audio}
        & 37.1 & 19.8 & 42.0
        & 43.8 & 20.5 & 45.2
        & 43.0 & 32.0 & 45.3
        & 25.6 & 11.5 & 30.3
        & 26.4 & 13.7 & 31.5 \\

        Audio Flamingo 3~\cite{audio_flamingo3}
        & 41.3 & 25.3 & 45.8
        & 46.8 & 28.9 & 49.4
        & 19.0 & 10.0 & 24.7
        & 34.6 & 28.1 & 39.8
        & 17.9 & 7.4 & 21.1 \\

        Qwen3-Omni~\cite{qwen3_omni}
        & 66.6 & 45.2 & 60.9
        & 72.8 & 52.0 & 65.1
        & 75.0 & 53.0 & 66.2
        & 47.1 & 38.5 & 49.6
        & 46.9 & 37.1 & 50.6 \\

        TimeAudio~\cite{timeaudio}
        & 75.3 & 60.3 & 67.2
        & 70.6 & 57.7 & 68.0
        & 18.0 & 8.0 & 23.0
        & 45.4 & 38.7 & 49.6
        & 30.3 & 12.0 & 29.6 \\

        SpotSound~\cite{spotsound}
        & 77.6 & 60.7 & 71.4
        & 58.6 & 42.6 & 56.0
        & \underline{81.0} & \underline{67.0} & 72.3
        & 48.5 & 39.0 & 51.0
        & \underline{91.8} & 85.5 & 86.0 \\

        \midrule

        \textbf{Ours} (Audio Flamingo 3)
        & \underline{84.4} & \underline{67.7} & \underline{76.1}
        & \textbf{85.2} & \underline{72.4} & \textbf{78.4}
        & 78.0 & 63.0 & \underline{72.7}
        & \underline{53.2} & \underline{41.9} & \underline{54.9}
        & 91.4 & \underline{86.7} & \underline{87.1} \\

        \textbf{Ours} (Qwen3-Omni)
        & \textbf{85.3} & \textbf{68.8} & \textbf{76.4}
        & \underline{85.0} & \textbf{72.8} & \underline{77.8}
        & \textbf{83.0} & \textbf{70.0} & \textbf{74.1}
        & \textbf{53.7} & \textbf{42.7} & \textbf{55.6}
        & \textbf{92.8} & \textbf{88.5} & \textbf{88.9} \\

        \bottomrule
    \end{tabular}%
    }
\end{table*}

\textbf{Frame-Adaptive Alignment and Fusion}.
Preserving $\widetilde{H}$ at the token level is important, since a natural language query may contain multiple semantic components, such as event identities and acoustic attributes, while a single audio frame may correspond to only a subset of them.
Aggregating $\widetilde{H}$ into a global representation would provide the same semantic condition to all frames and may obscure fine-grained correspondences between acoustic content and individual tokens~\cite{cgi}.
We therefore leverage cross-attention to allow each audio frame to adaptively retrieve relevant semantic information from the token-level sequence $\widetilde{H}$.
Specifically, the frame-level audio representations $X$ serve as the queries, while $\widetilde{H}$ provides the keys and values:
\begin{equation}
G
=
\operatorname{MHA}\!\left(
X,\widetilde{H},\widetilde{H}
\right)
=
\{g_t\}_{t=1}^{T}
\in \mathbb{R}^{T\times d},
\end{equation}
where $\operatorname{MHA}$ denotes multi-head attention, and $G$ denotes the resulting frame-level grounding representations.

To retain both representations while further modeling their interactions, the audio and grounding representations are fused at each frame as $z_t=\phi([x_t;g_t;x_t\odot g_t])$, forming the fused representations $Z=\{z_t\}_{t=1}^{T}\in\mathbb{R}^{T\times d}$, where $\odot$ denotes the element-wise product, $[\cdot;\cdot]$ denotes feature concatenation, and $\phi(\cdot)$ denotes the MLP used for feature fusion.
The fused sequence $Z$ is then processed by a temporal decoder to model dependencies across frames.
The decoder outputs are mapped by a linear head to frame-level localization scores, which are subsequently converted into temporal intervals.

\subsection{Temporal Reasoning}
\label{sec:reasoning}

Beyond temporal grounding, we further explore whether the trained grounding model can provide temporal evidence for downstream audio reasoning.
To this end, we use the grounding model as an external tool that the LALM can invoke during inference.
Specifically, the LALM first identifies the temporal operation required by the question, which guides its subsequent use of the grounding model.
When temporal evidence is useful, it formulates one or more concrete event queries and invokes the grounding model accordingly.
For each tool call, the generated event queries are provided together with the audio to the LALM, from which the corresponding textual hidden states are extracted.
These hidden states then condition the grounding model to predict the temporal intervals of the queried events.
The resulting intervals are organized into a structured tool output together with derived temporal attributes, including duration, first onset, and last offset.
The tool output is then incorporated into the LALM context as explicit temporal evidence for subsequent reasoning.
If further evidence is required, the LALM can formulate additional event queries and invoke the grounding model again before producing the final answer.

\section{Experiments}
\label{sec:experiment}

\subsection{Experimental Setup}
\label{sec:setup}

\textbf{Datasets}.
We conduct temporal grounding experiments on AudioGrounding (AG)~\cite{audiogrounding}, DESED~\cite{desed}, UnAV-100~\cite{unav100}, TACOS~\cite{tacos}, Clotho-Moment (Clotho-M)~\cite{am_detr}, and FTAR~\cite{timeaudio}.
This task aims to localize one or more temporal intervals of a target sound event in an audio recording.
The query is provided as a natural-language event description, while event labels directly serve as queries in label-based datasets.
For training, the corresponding training splits are used, with DESED restricted to real recordings and only the temporal audio grounding portion of FTAR included.
For evaluation, we use the test sets of AG, TACOS, and Clotho-M, the real test set of DESED, and the UnAV-100 subset~\cite{am_detr}.
The overall training and evaluation sets contain 98.1k and 14.5k samples, respectively.

\textbf{Implementation Details}.
We use Audio Flamingo 3~\cite{audio_flamingo3} and Qwen3-Omni~\cite{qwen3_omni} as the LALM backbones and ATST-Frame~\cite{atst_frame} as the audio encoder.
The shared feature dimension is set to 768, and the audio encoder features are upsampled to 50 Hz, with a four-layer CNN branch providing fine-grained acoustic details.
The fusion weight $w$ between the upsampled encoder features and CNN features is initialized to 0.5.
The cross-attention uses 12 attention heads, and the temporal decoder consists of five Conformer~\cite{conformer} layers with a hidden dimension of 768.
The grounding model is trained using frame-level binary cross-entropy loss, with binary frame labels derived from the annotated temporal intervals.
During training, the LALM and ATST-Frame remain frozen, while the remaining parameters are optimized for 20 epochs using AdamW with a learning rate of $5\times10^{-5}$ and a batch size of 32.
At inference, the frame-level localization scores are smoothed with a 15-frame median filter and thresholded at 0.3, after which consecutive positive frames are merged into temporal intervals.
We evaluate temporal grounding using IoU-based metrics.

\subsection{Main Grounding Results}
\label{sec:main_result}

Table~\ref{tab:main_results} presents the main temporal grounding results.
All baselines are evaluated on the same test sets using their publicly released checkpoints.
For FineLAP and DASM, longer recordings are split into non-overlapping 10s windows to match their input length.
The strong results of TimeAudio and SpotSound confirm that task-specific post-training can effectively improve the temporal localization capability of LALMs.
Our framework further improves performance with both LALM backbones, particularly at the stricter R@.7 threshold, with gains of 15.1 points on DESED and 8.1 points on AG over the best LALM baselines.
The proposed framework also performs strongly against specialized grounding models, with pronounced gains on TACOS and Clotho-M, which involve richer natural language queries, achieving 55.6 versus 46.7 and 88.9 versus 79.5 mIoU, respectively.
Overall, our framework consistently improves mIoU across all benchmarks, demonstrating the effectiveness of augmenting LALMs with frame-level grounding.

\begin{table}[!t]
    \centering
    \caption{
    Ablation study of key design choices in our framework.
    Results are reported in mIoU.
    Qwen3-Omni$^\dagger$ refers to audio representations from its native audio encoder.
    Bold variants indicate the settings used for the main results.
    }
    \label{tab:ablation}

    \renewcommand{\arraystretch}{1.35}
    \setlength{\heavyrulewidth}{0.06em}
    \setlength{\lightrulewidth}{0.035em}
    \setlength{\cmidrulewidth}{0.025em}
    \addtolength{\tabcolsep}{-1.5pt}

    \resizebox{0.95\columnwidth}{!}{%
    \begin{tabular}{@{}lccccc@{}}
        \toprule
        \textbf{Variant}
        & \textbf{AG}
        & \textbf{DESED}
        & \textbf{UnAV}
        & \textbf{TACOS}
        & \textbf{Clotho-M} \\
        \midrule

        \rowcolor[gray]{.97}
        \multicolumn{6}{c}{\textit{Query Representation}} \\

        MGA-CLAP~\cite{mga_clap}
        & 74.9 & 75.7 & 71.9 & 50.2 & 85.6 \\

        \begin{tabular}[c]{@{}l@{}}
            Qwen3-Omni \\[-3.6pt]
            (w/o audio context)
        \end{tabular}
        & 75.3 & 74.4 & 72.4 & 53.7 & 86.9 \\

        \textbf{Qwen3-Omni}
        & 76.4 & 77.8 & 74.1 & 55.6 & 88.9 \\

        \midrule

        \rowcolor[gray]{.97}
        \multicolumn{6}{c}{\textit{Audio Representation}} \\

        Qwen3-Omni$^\dagger$
        & 73.9 & 75.9 & 72.2 & 52.7 & 87.8 \\

        \begin{tabular}[c]{@{}l@{}}
            ATST-Frame \\[-3.6pt]
            (w/o CNN branch)
        \end{tabular}
        & 75.2 & 76.5 & 74.5 & 54.3 & 88.1 \\

        \textbf{ATST-Frame}
        & 76.4 & 77.8 & 74.1 & 55.6 & 88.9 \\

        \midrule

        \rowcolor[gray]{.97}
        \multicolumn{6}{c}{\textit{Query-to-Audio Alignment}} \\

        Global Pooling
        & 75.0 & 76.8 & 73.8 & 54.0 & 87.3 \\

        \textbf{Cross-Attention}
        & 76.4 & 77.8 & 74.1 & 55.6 & 88.9 \\

        \bottomrule
    \end{tabular}%
    }
\end{table}

\subsection{Ablation Studies}
\label{sec:ablation}

Table~\ref{tab:ablation} summarizes the ablation results using Qwen3-Omni as the LALM backbone.
To examine the role of LALM-based query modeling in our framework, we compare alternative query representations for conditioning the grounding model while keeping the remaining setup unchanged.
MGA-CLAP serves as a standalone query encoder in place of the LALM, as it is pretrained via audio-text contrastive learning and provides word-level features~\cite{mga_clap}.
To isolate the contribution of audio contextualization within the LALM, we further remove the audio input to Qwen3-Omni while retaining the same textual input.
The results show that the w/o audio context variant remains competitive with MGA-CLAP, while introducing audio context further improves performance.
This highlights the value of audio context in refining LALM-based query representations for temporal grounding.

We then investigate the effect of audio representations in our framework.
Using the native audio encoder of Qwen3-Omni in place of ATST-Frame leads to consistently lower performance across the benchmarks, indicating that representations from a specialized audio encoder are better suited to the fine temporal discrimination required for precise localization.
We further evaluate the contribution of the CNN branch by removing it from the audio representation.
ATST-Frame alone already provides strong performance, and incorporating the CNN features brings further improvements overall.
These results support the use of a specialized audio encoder, with the CNN branch contributing complementary fine-grained acoustic information.

Further, we evaluate the importance of aligning the token-level hidden states with audio representations at the frame level.
As a global alternative, we mean-pool the textual hidden states into a single query representation that is shared across all audio frames in the subsequent fusion.
In contrast, cross-attention performs frame-level alignment between hidden states and audio representations.
The consistent advantage of cross-attention over global pooling indicates that temporal grounding benefits from frame-specific alignment with token-level query representations rather than using a single global semantic condition across all frames.

\subsection{Temporal Reasoning}
\label{sec:reasoning_result}

We evaluate downstream temporal reasoning on the development split of the temporal soundscapes QA subset in DCASE 2025 Task 5~\cite{dcase_aqa}.
This benchmark contains multiple-choice questions designed to assess audio temporal reasoning.
For analysis, we group the questions according to the required temporal operation into Onset/Offset, Duration, Ordering, Counting, and Other categories.
We compare direct Qwen3-Omni inference with the same LALM using the trained grounding model as an external tool.

As shown in Table~\ref{tab:reasoning}, incorporating the grounding model improves the overall accuracy from 57.80\% to 71.26\%.
The largest improvements are observed for Duration and Onset/Offset, with gains of 32.08 and 29.17 percentage points, since these questions depend on explicit temporal information such as event boundaries or durations, which can be obtained directly from the grounding results.
For Ordering questions, whose answers are determined by the occurrence order of sound events, Qwen3-Omni already achieves relatively strong performance, while the grounding model provides a modest improvement.
The gains for Counting are also limited, as the frame-level localization objective does not explicitly supervise the number of event occurrences, and merged or fragmented intervals can lead to inaccurate counts.
Overall, these results highlight the value of frame-level grounding in providing temporal evidence for downstream reasoning.

\begin{table}[!t]
    \centering
    \caption{
    Temporal reasoning performance of direct LALM inference and using the grounding model as an external tool.
    \#Q denotes the number of questions in each category.
    The better result in each category is highlighted in bold.
    }
    \label{tab:reasoning}
    \renewcommand\arraystretch{1.5}
    \setlength{\heavyrulewidth}{0.06em}
    \setlength{\lightrulewidth}{0.035em}
    \setlength{\cmidrulewidth}{0.025em}
    \addtolength\tabcolsep{6pt}

    \resizebox{0.9\columnwidth}{!}{%
    \begin{tabular}{@{}lc>{\hspace{10pt}}c<{\hspace{-10pt}}c@{}}
        \toprule
        \multirow{2}{*}{\textbf{Question Type}}
        & \multirow{2}{*}{\textbf{\#Q}}
        & \multicolumn{2}{c}{\hspace{5pt}\textbf{Accuracy (\%)}\hspace{-5pt}} \\
        \cmidrule(l{15pt}r{5pt}){3-4}
        &
        & Direct
        & w/ Grounding Model \\
        \midrule

        Onset/Offset
        & 144 & 53.47 & \textbf{82.64} \\

        Duration
        & 106 & 50.94 & \textbf{83.02} \\

        Ordering
        & 151 & 71.52 & \textbf{74.17} \\

        Counting
        & 114 & 43.86 & \textbf{45.61} \\

        Other
        & 94 & 67.02 & 67.02 \\

        \midrule
        Overall
        & 609 & 57.80 & \textbf{71.26} \\

        \bottomrule
    \end{tabular}%
    }
\end{table}

\section{Conclusion}
\label{sec:conclusion}

In this paper, we presented a framework that augments LALMs with frame-level grounding to strengthen their fine-grained temporal perception.
Our framework leverages semantic representations from a frozen LALM to encode the event query with audio as context and combines them with fine-grained audio features in a dedicated grounding model for direct frame-level localization.
Extensive experiments demonstrate strong and consistent improvements across diverse grounding benchmarks, with particularly pronounced gains under stricter localization criteria.
Further, improvements in downstream temporal reasoning indicate that fine-grained localization can provide explicit temporal evidence for subsequent reasoning.
These findings suggest that frame-level grounding offers a complementary route toward fine-grained temporal perception in LALMs.

\clearpage

\bibliographystyle{IEEEbib}
\bibliography{refs}

\end{document}